\documentclass{iopart}
\usepackage{graphicx}

\begin{document}

\title{Negative Differential Mobility and Trapping in Active Matter Systems}
 
\author{C. Reichhardt and C. J. O. Reichhardt}
\address{Theoretical Division and Center for Nonlinear Studies,
Los Alamos National Laboratory, Los Alamos, New Mexico 87545, USA 
}
\ead{cjrx@lanl.gov}

\begin{abstract}
  Using simulations, we examine the average velocity as
  a function of applied drift force for active matter particles
  moving through a random obstacle array.
  We find that for low drift force,
  there is an initial
  flow regime where the mobility increases linearly with 
drive, while for higher drift forces a regime of negative differential mobility appears
in which the velocity decreases with
increasing drive due to the trapping of active particles behind obstacles.
A fully clogged regime exists at very high drift forces when
all the particles are permanently trapped behind obstacles.
We find for increasing
activity that the overall mobility is nonmonotonic, with an enhancement of
the mobility for small levels of
activity and a decrease in mobility for large activity levels.
We show how these effects evolve as a function of
disk and obstacle density, active run length, 
drift force, and motor force.
\end{abstract}

\maketitle

\section{Introduction}
There has been increasing interest in self-driven or active matter systems, which
are often modeled as a collection
of self-mobile disks with
mobility represented by
driven diffusion
or run-and-tumble dynamics \cite{1,2}.
When disk-disk interactions 
are included,        
an activity-induced clustering or
phase separation into a dense solid phase coexisting
with a low density active gas phase
can occur for sufficiently high activity and disk density.
Such clustering effects occur for both 
driven diffusive \cite{3,N,4,5,6,7} and run-and-tumble systems \cite{3,6,7,8}. 
Studies of active matter systems generally focus on samples with
featureless substrates, but recent work has addressed the
behavior
of active matter interacting with
more complex environments \cite{2}, such as 
random \cite{11,12,13} or periodic obstacle arrays \cite{14,15,16}, 
pinning arrays or rough landscape substrates \cite{16,17,18}, or funnel arrays \cite{N1},
as well as
mixtures of active and passive 
particles \cite{19}.
In run-and-tumble
disk systems, studies of
the average flux through an  
obstacle array in the presence of
an additional external drift force  \cite{11} 
show that for low activity or short run times,
the active disks have Brownian characteristics 
and are easily trapped; however, for  
increasing run persistence length or
activity, the trapping is reduced 
and the flux of disks through the system increases.
Interestingly, when the run time or activity is large,
the disk drift mobility 
is strongly reduced due to
enhanced self-trapping of disks behind the obstacles and by each other.  
Other studies of active particles moving 
through an array of obstacles
show that for increasing propulsion speed,
the particles
remain trapped behind obstacles for a longer time,
and as a result the long time diffusion constant is decreased
for high activity particles compared to passive or 
Brownian particles \cite{13}. 
If the obstacles are replaced by a rough substrate, another
study showed
that the drift mobility of the
particles increases
with increasing run length
since the self-clustering effect allows the particles
to act like an effective larger-scale rigid object that couples only weakly
to the substrate
\cite{18}.
Studies of flocking active particles moving
through obstacle arrays
reveal nonmonotonic behavior as a function of disorder strength \cite{20} 
and a disorder-induced flocking to non-flocking transition \cite{12,21}.

Here we examine run-and-tumble active matter disks in the presence of 
a random array of obstacles where we apply an external driving force in order to
measure 
the long time average disk drift velocity
in the direction of drive. 
While in previous work we considered a constant
applied drift force \cite{11}, here we 
examine the effects of varied drift forces and compare
the resulting velocity-force curves to
those found in other systems
that exhibit depinning, such as
passive particles driven over random disorder \cite{N2}.
Previous studies
of passive or Brownian particles driven though
an obstacle array
showed that there 
can be a regime of   
negative differential mobility
where
the velocity {\it decreases} 
with increasing drive, and
the velocity can even drop to zero in the limit of large drive
\cite{22,23,24}.
Negative differential mobility also appears
 when the obstacles themselves are allowed to move
 \cite{25,26,27}.
 Such effects can arise for 
driven particles in  laminar flows \cite{28}, particles driven through glass formers \cite{29},
vortices in type-II superconductors moving   
through periodic pinning arrays\cite{N2,30,31}, colloids moving
on ordered pinning arrays \cite{33},
and in nonequilibrium states of certain types of semiconductors \cite{34}. 

We specifically examine run-and-tumble driven disks, where a motor force
$ F_m^i{\bf \hat m}$  acts on disk $i$ during a fixed running time $\tau$ in a randomly
chosen direction ${\bf \hat m}$.  At the end of the running time,
a new run begins with the motor force acting in a new randomly chosen direction.
The disks move through a random array of obstacles under
an external drive $F_{D}{\bf \hat x}$,
and we measure the average disk drift velocity
$\langle V\rangle$
in the driving direction as a function of increasing $F_{D}$.
At low drives 
$F_{D} \ll F_{M}$,
the drift velocity $\langle V\rangle$ increases
linearly with increasing $F_{D}$,
but we find that $\langle V\rangle$ 
reaches a maximum and then
decreases with increasing $F_{D}$
due to the partial trapping of disks behind obstacles.
For large enough drives $F_D>F_{cl}$  we observe
a fully clogged state with $\langle V\rangle = 0$.
The value of $F_{cl}$
increases with increasing activity and saturates for long run times, 
while the maximum value of $\langle V\rangle$ changes
nonmonotonically as a function of run time.
In some cases, a system with longer run times has lower mobility for small $F_D$ but
higher mobility at large $F_D$ compared to
a system with shorter run times.
For fixed run time and increasing motor force $F_m$,
we find that the transition from a linear dependence of $\langle V\rangle$ on $F_m$
to a decrease in $\langle V\rangle$ with increasing $F_m$ follows the line
$F_m=F_D$.
When we increase the disk density,
we observe crowding effects that reduce the overall mobility but can also increase the
range of parameters for which
negative differential mobility occurs.   

\section{Simulation}
We consider a two-dimensional system
of size $L \times L$ with periodic boundary conditions in the
$x$ and $y$ directions, where $L=50$. 
Within the system we place  $N_a$ active disks of radius $R=0.5$ and
$N_{\rm obs}$ obstacles which are modeled as stationary disks of the same size as
the active disks.
The area coverage of the obstacles is
$\phi_{\rm obs} = N_{\rm obs}\pi R^2/L^2$, the
area coverage of the active disks is $\phi_{a} = N_{a}\pi R^2/L^2$,
and the total area coverage is $\phi_{\rm tot} = \phi_{\rm obs} + \phi_{a}$.    
The disk-disk interactions are modeled as a short range
harmonic repulsion
${\bf F}_{dd} = k(d - 2R)\Theta(d-2R){\hat {\bf d}}$ 
where $d$ is the distance between the disks, ${\hat {\bf d}}$ is the displacement 
vector between the disks, $k$ is the spring constant,
and $\Theta$ is the Heaviside step function.
The dynamics of an active disk $i$ are governed by
the following overdamped
equation of motion: $\eta d{\bf r}_i/dt = {\bf F}_{aa}^i + 
{\bf F}_{m}^i + {\bf F}_{\rm obs}^i + {\bf F}_{D}$ where 
the damping constant $\eta = 1.0$.
Here ${\bf F}_{aa}^i=\sum_{j\ne i}^{N_a}F_{dd}^{ij}$ is the interaction between
active disks and ${\bf F}_{\rm obs}^i=\sum_{k}^{N_{\rm obs}}F_{dd}^{ik}$ is the
interaction between active disk $i$ and the obstacles.
Each mobile disk has a motor force
${\bf F}_{m}^i=F_m{\bf \hat m}^i$ applied in a randomly chosen direction
${\bf \hat m}^i$ 
which changes after each run time $\tau$.
In the absence of any collisions, during a running time the motor force 
translates the disk
a distance $l_{r} = F_{m}\tau \delta t$, where $\delta t = 0.002$ is the simulation
time step.
The mobile disks experience an additional external driving force in the $x$ direction,
${\bf F}_{D}=F_D{\bf \hat x}$.
After changing
$F_{D}$, we wait for at least $10^8$ simulation time steps to ensure that
we have reached a steady state flux before measuring the
average drift velocity 
in the direction of drive,
$\langle V\rangle = N_{a}^{-1}\sum^{N_a}_{i}\langle {\bf v \cdot {\bf \hat x}}\rangle_{i}$.
We quantify the activity level of the disks
in terms of $l_{r}$ and $F_{m}$.  
      
\begin{figure}
  \center
\includegraphics[width=0.7\columnwidth]{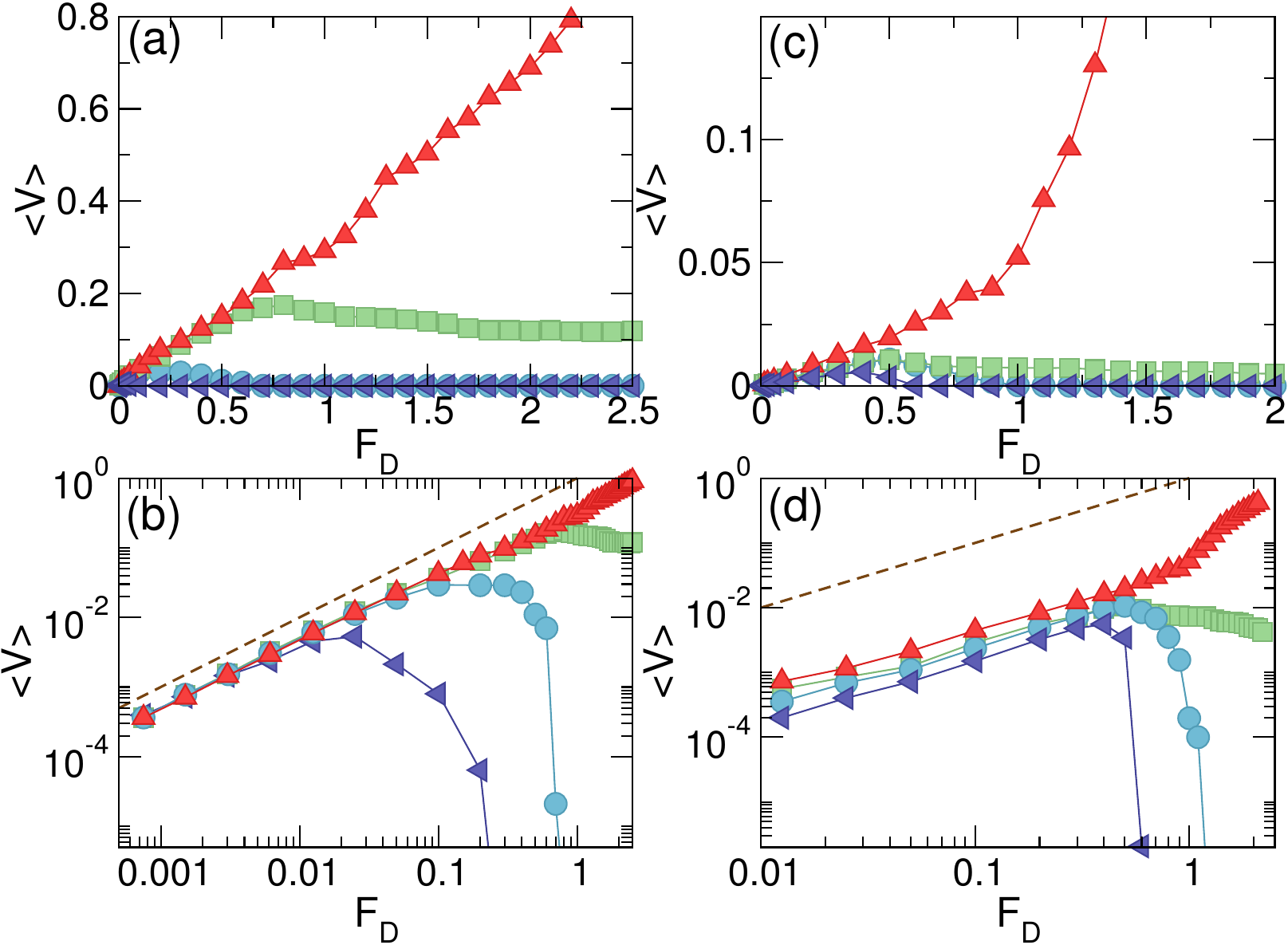}
\caption{ Average drift velocity $\langle V\rangle$ vs $F_{D}$
  in samples with $\phi_{\rm obs} = 0.1257$ and $F_m=0.5$ for
  $\phi_{a} = 0.00785$ (dark blue left triangles),
  $0.03146$ (light blue circles), 
$0.0628$ (green squares), and $0.1257$ (red up triangles).
  (a) $l_r=0.01$. (b) Data from (a) on a log-log scale highlighting the
  negative differential mobility for
  $\phi = 0.00785,$ 0.03146, and 0.0628. 
  (c) $l_{r} =  120$. (d) Data from (c) on a log-log scale.
  The dashed lines in panels (b) and (d) indicate the linear
  behavior $\langle V\rangle=F_D$ in an obstacle-free system.
}
\label{fig:1}
\end{figure}

\section{Results}

In Fig.~\ref{fig:1}(a) we plot $\langle V\rangle$ versus
$F_{D}$ for a system with $F_{m} = 0.5$, $\phi_{\rm obs}=0.1257$, and
$l_{r} = 0.1$
at $\phi_{a} = 0.00785$, 0.03146, 0.0628, and $0.1257$,
and in Fig.~\ref{fig:1}(b) we show the same
data on a log-log scale.
At the low density of $\phi_{a} = 0.00785$,
the sample behavior falls in the
single active disk limit and
$\langle V\rangle$ increases linearly 
with $F_{D}$
before reaching a maximum near
$F_{D} = 0.03$ and then decreasing
to  $\langle V\rangle = 0.0$
for $F_{D} \geq 0.21$.
For $\phi_{a} = 0.03146$, $\langle V\rangle$
reaches a  maximum near $F_{D} = 0.3$ 
and drops to zero for $F_{D} \geq 0.71$. 
In both cases the system exhibits 
what is called negative differential mobility (NDM),
where the average velocity decreases with
increasing $F_{D}$,
and at high enough drives the sample
reaches a pinned or clogged state.  
For $\phi = 0.0628$  there is still a region of NDM for $F_{D} > 0.8$;
however, the velocity remains finite up to the
maximum drive $F_D/F_m = 5$ that  we consider.
For $\phi = 0.1257$,
when the number of disks equals the number of obstacles, the velocity
monotonically increases 
with $F_{D}$ and the NDM is lost.
In Fig.~\ref{fig:1}(b) the dashed line indicates the linear behavior
$\langle V\rangle = F_{D}$ of an obstacle-free system.
For all values of $\phi_a$, we find the same linear increase
of $\langle V\rangle$ with $F_D$ for small $F_D$, 
and the value of $F_{D}$ at which NDM appears shifts to
higher values of
$F_{D}$ as $\phi_a$ increases.

\begin{figure}
  \center
\includegraphics[width=0.8\columnwidth]{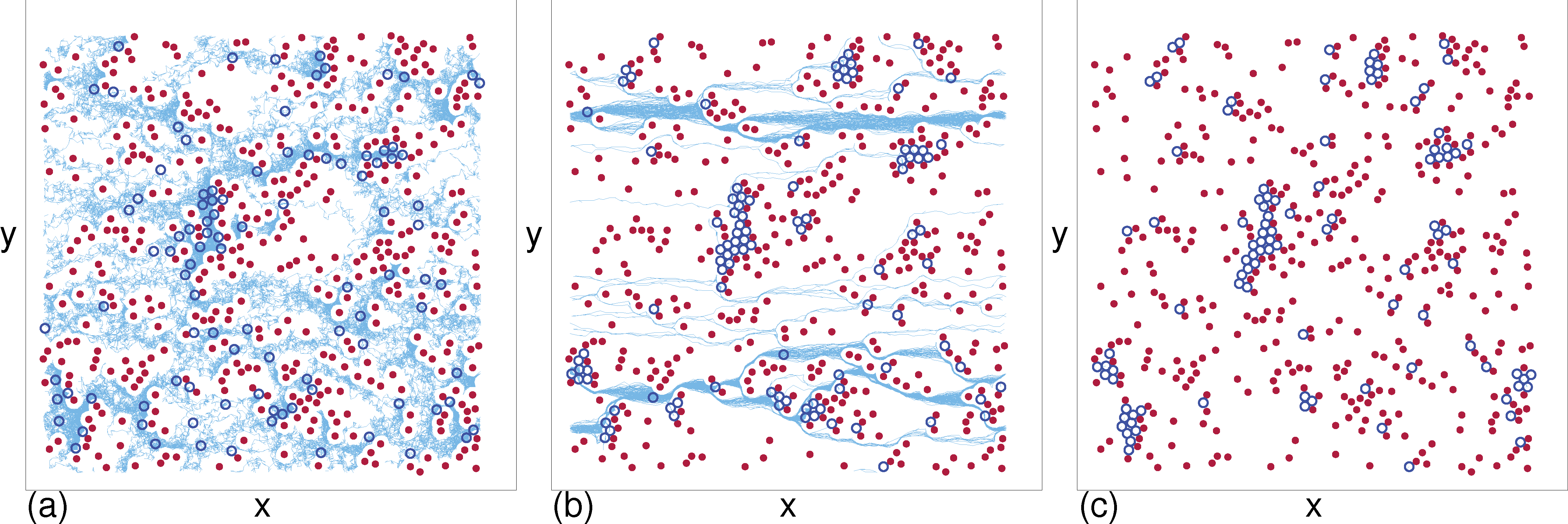}
\caption{ The obstacle positions (red filled circles),
  active disks (dark blue open circles),
  and trajectories (light blue lines) for the
  system from Fig.~\ref{fig:1}(a,b) 
with $\phi_{\rm obs} = 0.1257$, 
$\phi_{a} = 0.03146$, and $l_{r} = 0.01$.
(a) $F_{D} = 0.0125$,  where $\langle V\rangle$ is increasing with
increasing $F_{D}$.
(b) $F_{D} = 0.6$, where partial trapping of active disks
behind the obstacles occurs and the system exhibits negative
differential mobility.
(c) $F_{D} = 1.0$, where $\langle V\rangle = 0$
and the system is in a completely clogged state.
}
\label{fig:2}
\end{figure}

In Fig.~\ref{fig:2} we show the trajectories of the active disks
at three different values of $F_D$ for the $\phi_a=0.03146$ system from   
Fig.~\ref{fig:1}(a,b).
At $F_D=0.0125$ in
Fig.~\ref{fig:2}(a),
$\langle V\rangle$ is increasing 
with increasing $F_{D}$.
The disk trajectories are space filling,
and $F_{D}/F_{m} = 0.025$ is small enough
that when a disk becomes trapped behind
an obstacle,
the motor force is large enough to move the disk in the direction opposite
to the drive, 
permitting the disk to 
work its way
around the obstacle.
As a result, the active disks can easily explore nearly all the 
possible paths through the obstacle array.
As $F_{D}$ increases, the
ability of the disks to back away from an obstacle
is reduced, and
the amount of time disks spend trapped behind obstacles  increases,
as illustrated in Fig.~\ref{fig:2}(b) at $F_{D} = 0.6$,
corresponding to $F_{D}/F_{m} = 1.2$, where the 
system exhibits NDM.
Here there are several locations in which the disks become
trapped for long periods of time.
As $F_{D}$ is further increased, more
disks become trapped and
$\langle V\rangle$ diminishes, as shown in
Fig.~\ref{fig:2}(c) for $F_{D} = 1.0$, where
all the active disks are permanently trapped
and $\langle V\rangle = 0$.
Two effects reduce the trapping susceptibility as $\phi_a$ increases.
Once a portion of the disks becomes trapped behind the
most confining obstacle configurations, additional active disks
can no longer be trapped at these same locations, meaning that the
``deepest'' traps are effectively inactivated.
In addition,
at locations
where $N$ multiple active disks are 
trapped one behind another, there is a
finite probability 
that the motor forces of these disks will simultaneously be oriented
in the direction opposite to that of the drift force, permitting the
disks to escape,
so that complete trapping
will occur only when
$F_{D} > N F_{m}$.

\begin{figure}
  \center
\includegraphics[width=0.8\columnwidth]{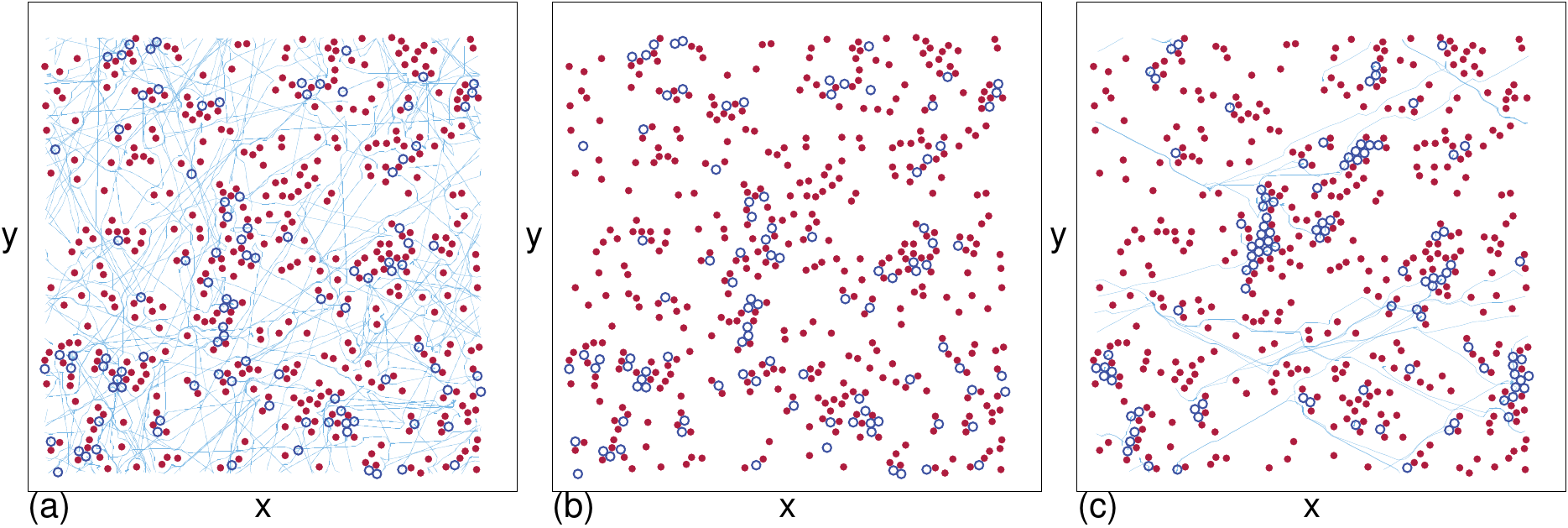}
\caption{The obstacle positions (red filed circles), active disks
  (dark blue open circles), 
and trajectories (light blue lines) for the system in Fig.~\ref{fig:1}(c,d)
with $\phi_{\rm obs} = 0.1257$, 
$\phi_{a} = 0.03146$, and $l_{r}= 120$. 
(a) $F_{D}= 0.0125$, where the disks move in straight lines.
(b) The same as in (a) but without the trajectories,
showing that nearly all of the active disks are in contact with an obstacle.
(c) $F_{D}= 1.0$,  where there is increased trapping
but $\langle V\rangle$ remains finite, unlike
the $l_{r} = 0.01$ case illustrated in Fig.~\ref{fig:2}(c) where complete trapping occurs.
}
\label{fig:3}
\end{figure}
 
In Fig.~\ref{fig:1}(c,d) we plot $\langle V\rangle$ versus $F_{D}$
for the same system in 
Fig.~\ref{fig:1}(a,b)
but with a much larger run length of $l_{r} = 120$.
We observe
the same general features 
as in the $l_{d} = 0.01$ case; 
however, the value of $F_{D}$
at which $\langle V\rangle$ reaches zero is shifted upward
and the magnitude of $\langle V\rangle$ for a given
$F_{D}$ is significantly reduced,
as indicated by comparing the curves in Fig.~\ref{fig:1}(d) to the dashed
line which is the flow expected in an obstacle-free system.

Figure~\ref{fig:3}(a) shows the active disk
trajectories
for the system in Fig.~\ref{fig:1}(c,d) with $\phi_a = 0.03146$ and $l_r=120$
at $F_{D} = 0.0125$.
In this regime, $\langle V\rangle$ is increasing with
increasing $F_{D}$;  however, $\langle V\rangle$ is
smaller by nearly a factor of
20 than in
the $l_{r} = 0.01$ case illustrated in Fig.~\ref{fig:2}(a).
The active disks
in Fig.~\ref{fig:3}(a) are not strongly affected by the external drive and
move in straight lines while running; 
however, upon encountering an obstacle the active disk
pushes against it and becomes self-trapped, reducing the mobility of the system.
To more clearly demonstrate the self-trapping effect
that occurs for large run lengths,
in Fig.~\ref{fig:3}(b) we plot
the same snapshot of the active disk and obstacle positions
without trajectories, and find that nearly all of the active disks are in contact
with an obstacle.
At $F_D=1.0$,
as
illustrated in
Fig.~\ref{fig:3}(c),
$\langle V\rangle$ is finite
in the $l_r=120$ system, whereas
$\langle V\rangle=0$ for $l_r=0.01$.
Here, although a
considerable amount of disk
trapping occurs,
the longer run times allow some of the disks
to become mobile, giving a nonzero contribution to $\langle V\rangle$.

\begin{figure}
  \center
  \includegraphics[width=0.7\columnwidth]{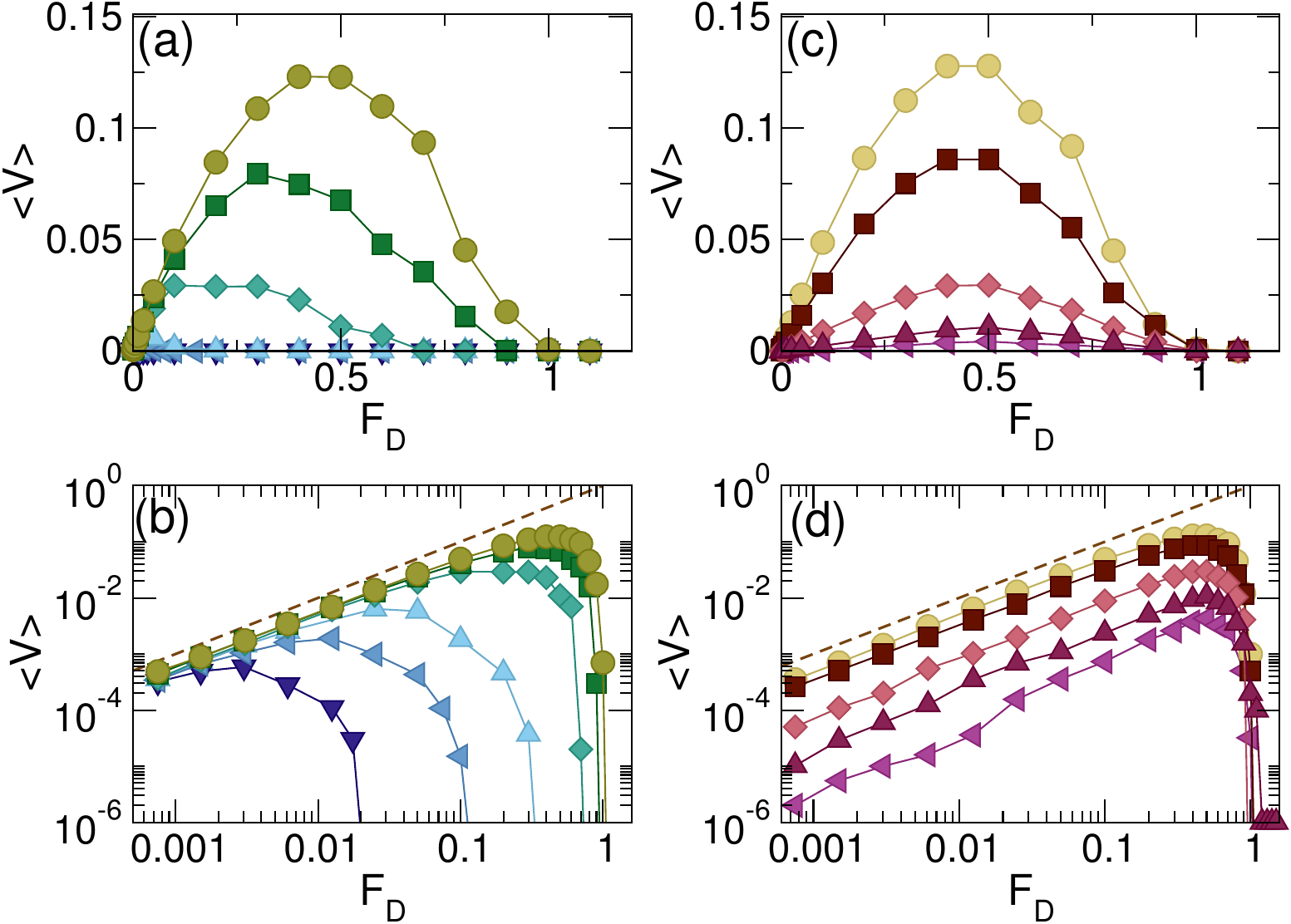}
  \caption{ $\langle V\rangle$ vs $F_{D}$
    in samples with $\phi_{\rm obs} = 0.1257$, $F_m=0.5$, and $\phi_{a} = 0.03146$. 
    (a) $l_{r} = 0.002$
    (dark blue down triangles),
    $0.01$
    (medium blue right triangles),
    $0.02$
    (light blue up triangles),
    $0.1$
    (teal diamonds),
    $0.3$
    (dark green squares),
    and $1.0$
    (light green circles),
    from bottom to top.
    (b) The curves from panel (a) plotted
    on a log-log scale.  The dashed line indicates
the
    obstacle-free limit of $\langle V\rangle = F_{D}$.
    Here $\langle V\rangle$ increases 
    with increasing $l_{r}$.
    (c) $\langle V\rangle$ vs $F_D$ for the same system with
    $l_{r} = 3$
    (yellow circles),
    $10$
    (dark red squares),
    $40$
    (light pink diamonds),
    $120$
    (dark pink up triangles),
    and $320$
    (magenta triangles),
    from top to bottom.
    (d) The curves from panel (a) plotted on a log-log scale.
    The dashed line indicates the obstacle-free limit of
    $\langle V\rangle = F_{D}$. Here  $\langle V\rangle$ decreases with 
increasing $l_{r}$.
}
\label{fig:4}
\end{figure}

To get a better understanding of
how $l_{r}$ affects the shape of the velocity-force
curves, in Fig.~\ref{fig:4}(a)
we plot $\langle V\rangle$ versus $F_{D}$
for a sample with
$\phi_{\rm obs} = 0.1257$, $\phi_{a} = 0.03146$, and $F_{m} = 0.5$
at $l_r$ values ranging from $l_r=0.002$ to $l_r=1.0$.
The same curves are shown on a log-log scale in 
Fig.~\ref{fig:4}(b),
where the dashed line indicates the
obstacle-free limit $\langle V\rangle = F_{D}$.
We note that for non-active disks with $l_{r} = 0$,
$\langle V\rangle = 0$ for all drive values at this disk density.
Three trends emerge from the data.
There is an overall increase in $\langle V\rangle$ with increasing $l_{r}$ 
for all values of $F_{D}$.
Additionally, both the maximum value of
$\langle V\rangle$ and the drive at which $\langle V\rangle$ reaches zero
shift to higher values of $F_{D}$ with increasing $l_r$.
Figure~\ref{fig:4}(c) shows $\langle V\rangle$ versus $F_D$ in the same
sample for $l_r$ values ranging from $l_r=3$ to $l_r=320$, and
in Fig.~\ref{fig:4}(d) we plot the same curves on a log-log scale.
Here there is an overall decrease in $\langle V\rangle$ with increasing
$l_r$ for all values of $F_D$.
The drive at which $\langle V\rangle$ reaches zero has its largest
value of $F_D=1.0$ for $l_r=1.0$ and decreases with increasing
$l_r$ for $l_r>1.0$.
In the limit $l_{r} \rightarrow \infty$,
$\langle V\rangle = 0$ for all $F_D$ since fluctuations in the disk motion
are eliminated, so 
the system cannot escape from a clogged state.  This is similar to what occurs in
the $l_r=0$ nonactive disk system.

\begin{figure}
  \center
\includegraphics[width=0.6\columnwidth]{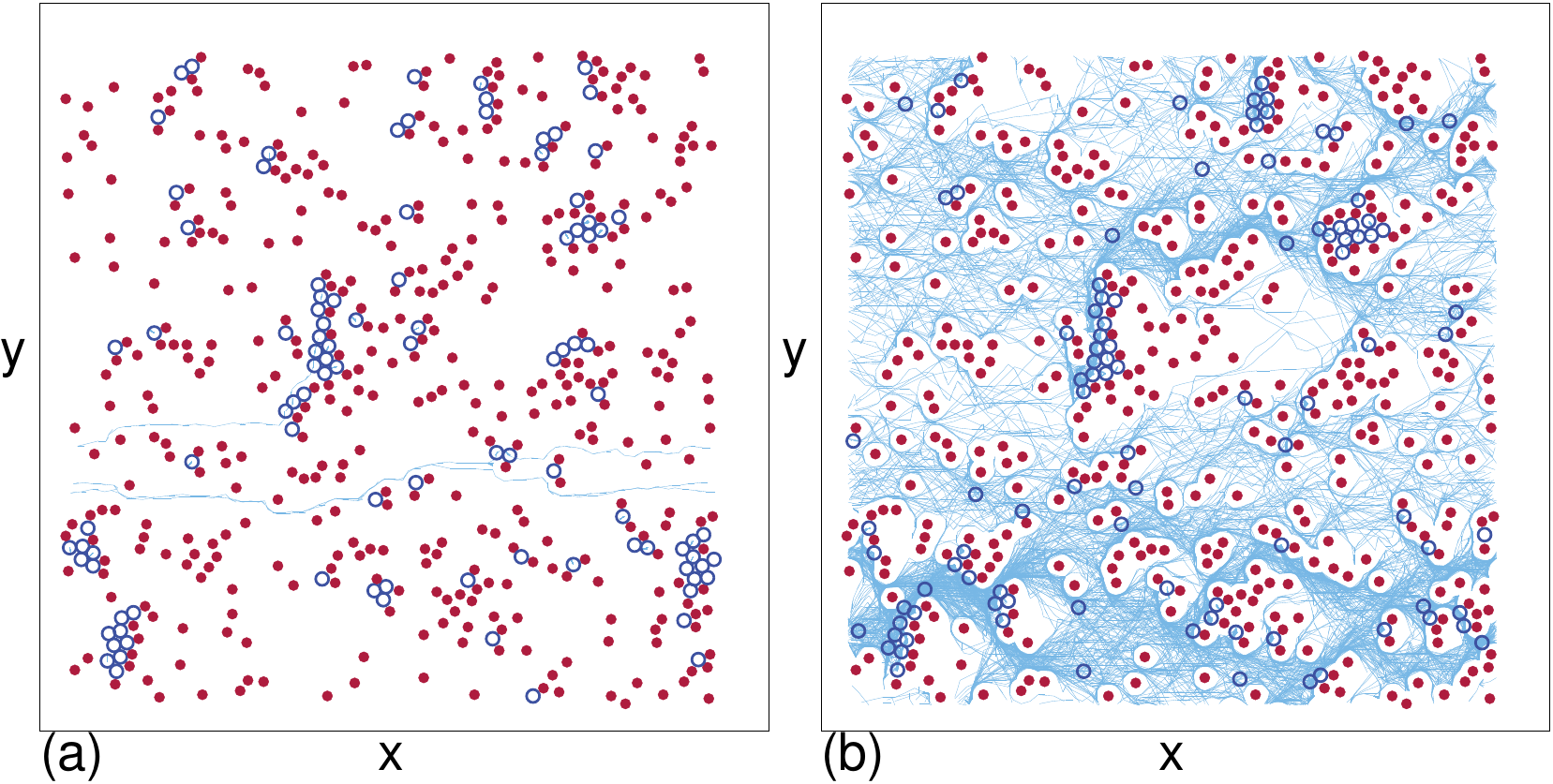}
\caption{ The obstacle positions (red filled circles), active disks (dark blue
  open circles), and trajectories (light blue lines)
  for the system in Fig.~\ref{fig:4}(a)
  at $\phi_{\rm obs} = 0.1257$, $F_m=0.5$, $\phi_a = 0.03146$, and $F_{D}= 0.5$. 
  (a) At $l_{r} = 0.02$, most disks are trapped.
  (b) At $l_{r} = 3.0$, the flow through the sample is optimized.
}
\label{fig:5}
\end{figure}

In Fig.~\ref{fig:5}(a) we illustrate the active disk trajectories for
the system in Fig.~\ref{fig:4} at $F_{D} = 0.5$
for $l_{r} = 0.02$, where most of the disks are trapped.
At later times, all the disks become trapped and $\langle V\rangle=0$.
In Fig.~\ref{fig:5}(b) we show the same system at $l_r=3.0$ and $F_D=0.5$,
where $\langle V\rangle$ passes through its maximum value in
Fig.~\ref{fig:4}(c,d).
Here
the disks can move easily through the system
and the amount of trapping is significantly reduced.
At higher values of $F_{D}$
more trapping occurs,
and for large enough $F_{D}$, $\langle V\rangle = 0$.

\begin{figure}
  \center
\includegraphics[width=0.7\columnwidth]{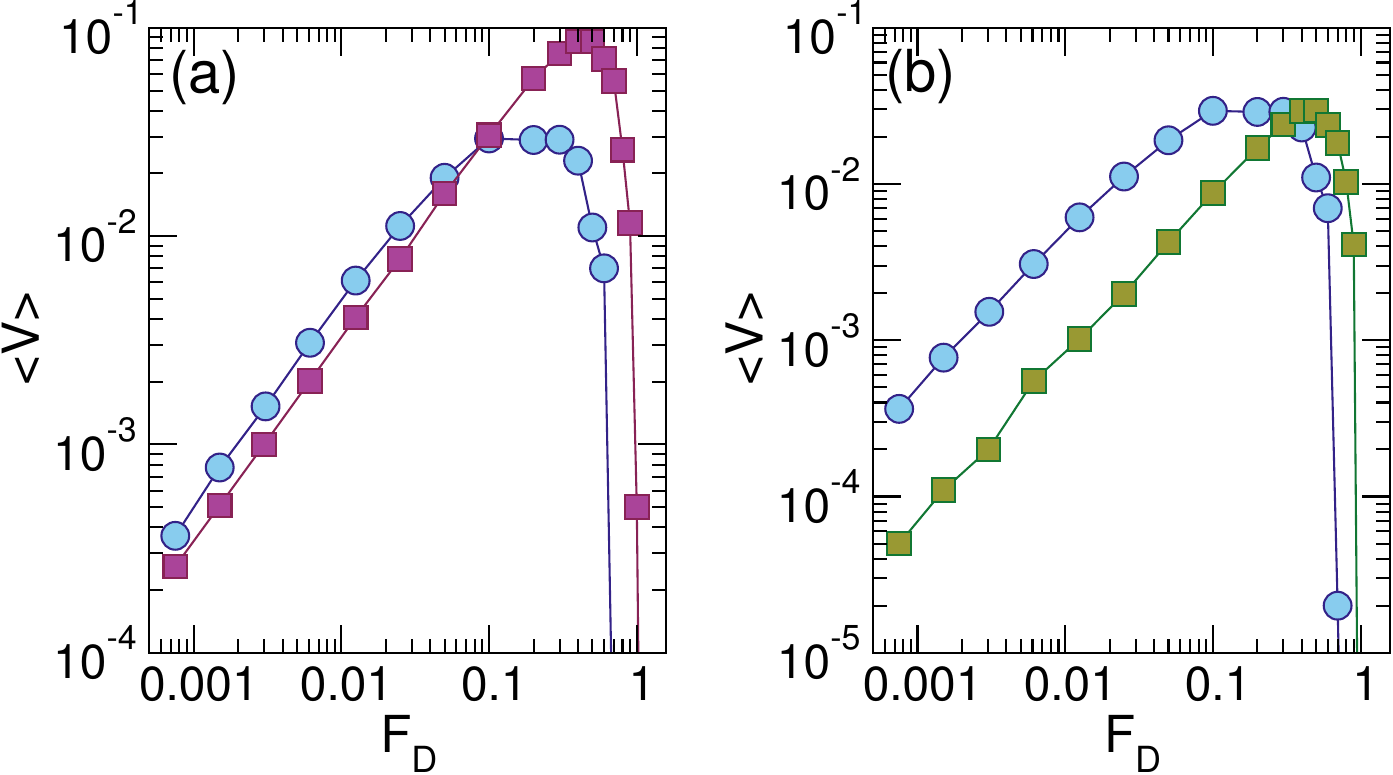}
\caption{ The drift velocity $\langle V\rangle$ vs $F_{D}$ for the system in 
Fig.~\ref{fig:3} with $\phi_{\rm obs} =  0.1257$, $F_m=0.5$, and $\phi_{a} = 0.03146$.
(a) $l_{r} = 0.1$ (blue circles) and $l_{r} = 10$ (magenta squares).
(b) $l_{r} = 0.1$ (blue circles) and $l_{r} = 40$ (green squares).  In both
cases there is a crossing of the curves.
}
\label{fig:6}
\end{figure}

\begin{figure}
  \center
\includegraphics[width=0.7\columnwidth]{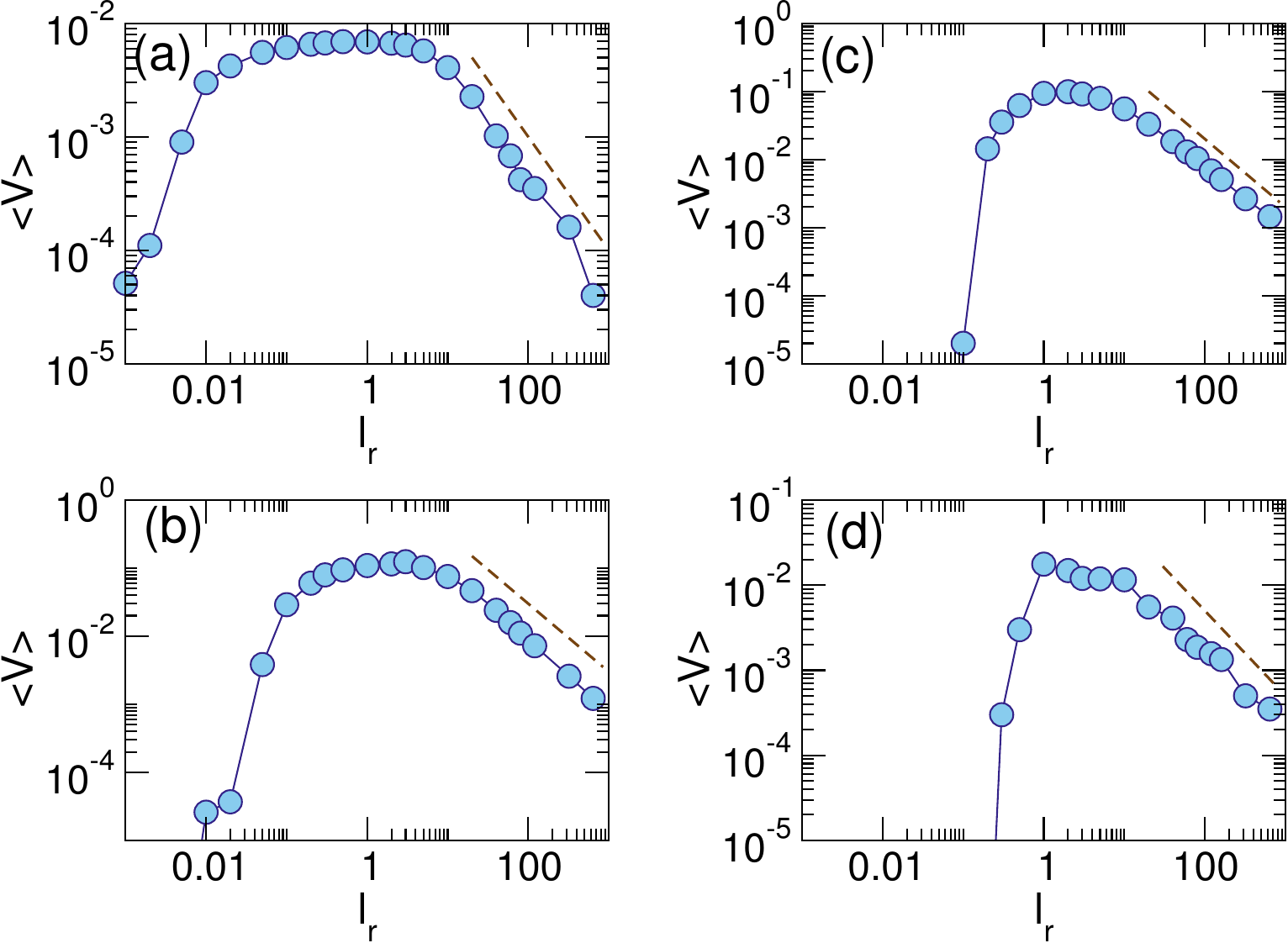}
\caption{ $\langle V\rangle$ vs $l_{r}$ for
  samples with $\phi_{\rm obs} = 0.1257$, $F_m=0.5$,  and $\phi_{a} = 0.03146$.
  (a) $F_{D} = 0.0125$.  (b) $F_{D} = 0.3$.
  (c) $F_{D} = 0.7$.  (d) $F_{D} = 0.9$.
  Dashed lines indicate fits to $\langle V\rangle \propto l^{-1}_{r}$
  }
\label{fig:7}
\end{figure}

\begin{figure}
  \center
\includegraphics[width=0.8\columnwidth]{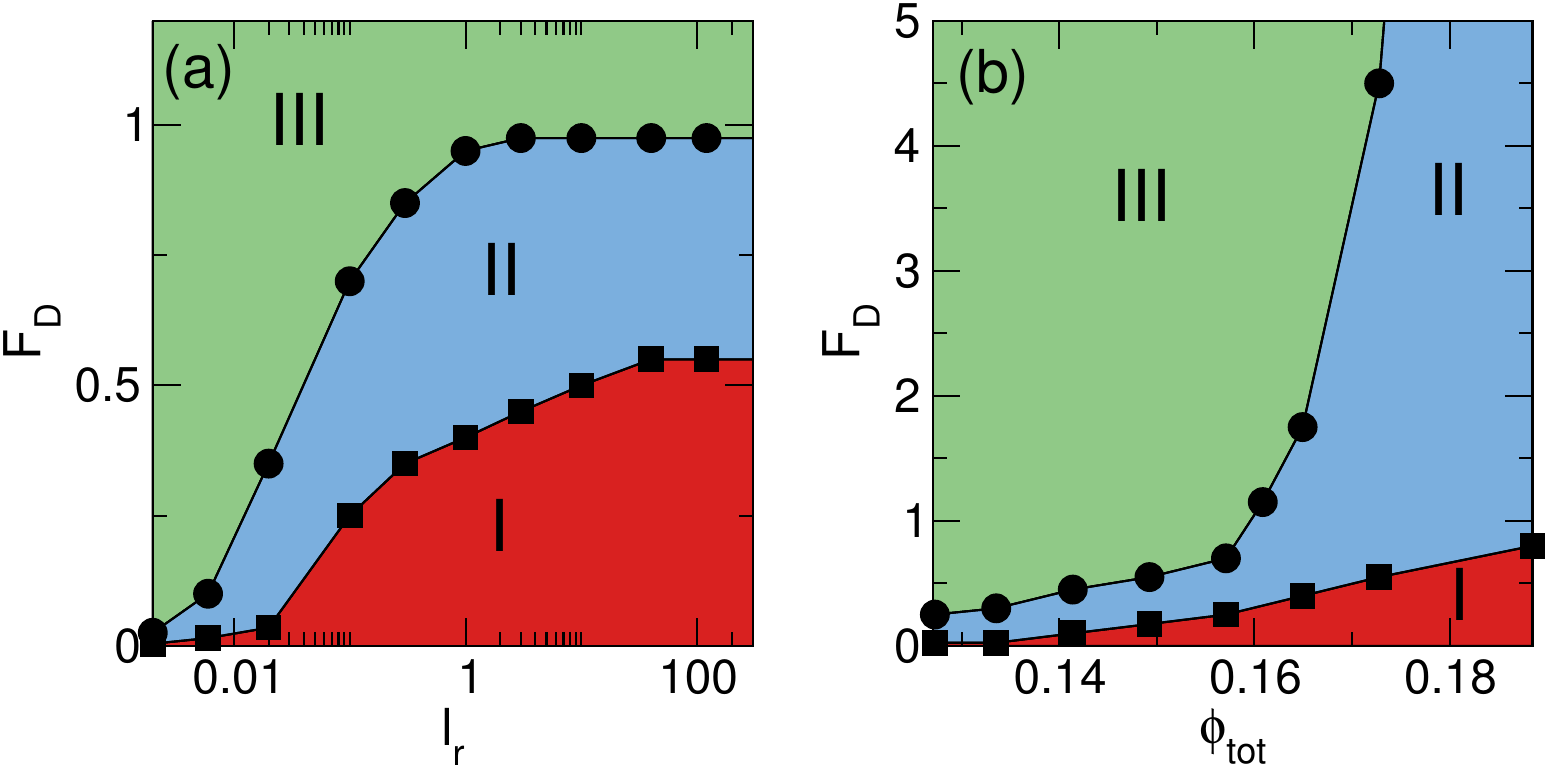}
\caption{Dynamic phase diagrams.  Phase I (red)
  is the ohmic flow regime in which $\langle V\rangle$ increases with
  increasing $F_D$.
  Phase II (blue) is the partial trapping regime where NDM occurs.
  Phase III (green) is the complete clogging regime
  in which $\langle V\rangle = 0$.
  (a) Dynamic phase diagram as a function of $F_{D}$ vs $l_{r}$ for a 
system with $\phi_{\rm obs} = 0.1257$, $\phi_{a} = 0.03146$,
and $F_{m} = 0.5$.
(b) Dynamic phase diagram as a function of $F_{D}$ vs
$\phi_{\rm tot}$ for $\phi_{\rm obs} = 0.1257$ and $F_{m} = 0.5$.
}
\label{fig:8}
\end{figure}

The value of $l_{r}$ that maximizes the flux through the system depends
strongly on $F_D$.
In Fig.~\ref{fig:6}(a) we plot $\langle V\rangle$ versus $F_{D}$ for 
the system in Fig.~\ref{fig:4}
at $l_{r} = 0.1$ and $l_{r} = 10$.
For $F_D<0.1$, $\langle V\rangle$ is larger in the $l_r=0.1$ system
than in the $l_r=10$ system, while for $F_D>0.1$, the situation is reversed
and $\langle V \rangle$ is largest in the $l_r=10$ system.
A comparison of $\langle V\rangle$ versus $F_D$ for $l_r=0.1$ and
$l_r=40$ appears in Fig.~\ref{fig:6}(b), 
where $\langle V \rangle$ is larger in
the $l_r=0.1$ system for $F_D < 0.3$
and larger in the $l_r=40$ system for $F_D>0.3$, while the maximum value
of $\langle V\rangle$ is nearly the same for both values of $l_r$.
This result has implications for active particle separation or mixing, and
indicates that a less active particle species would move faster under
a drift force at smaller $F_D$ than a more active particle species.
At larger $F_D$ the reverse would occur, with the less active species
becoming immobile while the more active particles are still able to
flow through the system.
The curves in Fig.~\ref{fig:6} also indicate
that is possible to tune $F_{D}$ such that active particle species with
very different activity levels
drift with equal values of $\langle V\rangle$,
such as by setting $F_D=0.3$ for the $l_r=0.1$ and $l_r=40$ disks in
Fig.~\ref{fig:6}(b). 
It is possible 
that certain living systems
such as bacteria subjected to an external drift
may actually lower their activity in order to move through a 
heterogeneous environment
if the external flow is weak,
while if there is a strong drift flow,
an increase in the activity level would
improve the mobility.

In Fig.~\ref{fig:7}
we plot $\langle V\rangle$ versus $l_{r}$ for the system in
Fig.~\ref{fig:1} with $\phi_{\rm obs} = 0.1257$, $F_m=0.5$, and $\phi_{a} = 0.03146$
for different values of $F_D$.
For $F_{D} = 0.0125$ in Fig.~\ref{fig:7}(a),
$\langle V\rangle$ initially
increases with increasing $l_{r}$ before reaching 
a maximum value at $l_{r} = 1.0$, after which it drops
by several orders of magnitude as $l_{r} \rightarrow 320$.
At $F_D=0.3$ in Fig.~\ref{fig:7}(b), $F_D=0.7$ in Fig.~\ref{fig:7}(c), and
$F_D=0.9$ in Fig.~\ref{fig:7}(d), 
$\langle V\rangle = 0$ at small values of $l_r$, and as $l_r$ increases
$\langle V\rangle$ passes through a maximum value before decreasing again.
In the larger $l_r$ regime where $\langle V\rangle$ is a decreasing
function of $l_r$, 
the drift velocity approximately follows the form
$\langle V\rangle \propto 1/l_{r}$, as indicated by the dashed
line fits in each panel.

In Fig.~\ref{fig:8}(a) we show the evolution of the
three different phases
as a function of $F_{D}$ versus $l_{r}$ 
for a system with $\phi_{\rm obs} = 0.1257$, $\phi_{a} = 0.03146$, and
$F_{m} = 0.5$.  In phase I, the ohmic flow regime,
$\langle V\rangle$ increases with
increasing $F_{D}$.
In phase II, partial trapping occurs and we observe NDM.
Phase III is the 
completely clogged regime with $\langle V\rangle = 0$.
The extent of phase I grows as $l_r$ increases until the I-II boundary saturates
at large $l_r$
to the value  $F_D=0.5$,
corresponding to $F_D/F_m=1.0$.
Similar behavior appears for phase II, with a saturation of the
II-III boundary for $l_r>1$
to $F_{D} = 1.0$.
The onset of phase III drops to
$F_{D} = 0$ when $l_{r} = 0$, indicating that for this density, nonactive disks are
permanently clogged.
In Fig.~\ref{fig:8}(b) we plot a dynamic phase diagram as a function of
$F_{D}$ versus $\phi_{\rm tot}$, where we vary $\phi_{\rm tot}$ by
fixing $\phi_{\rm obs} = 0.1257$ and changing $\phi_{a}$.
Phase III disappears
for $\phi_{\rm tot} > 0.17$, and the extent of
phase I increases
as the ratio $\phi_a/\phi_{\rm obs}$ of active disks to obstacles increases.

\begin{figure}
  \center
\includegraphics[width=0.7\columnwidth]{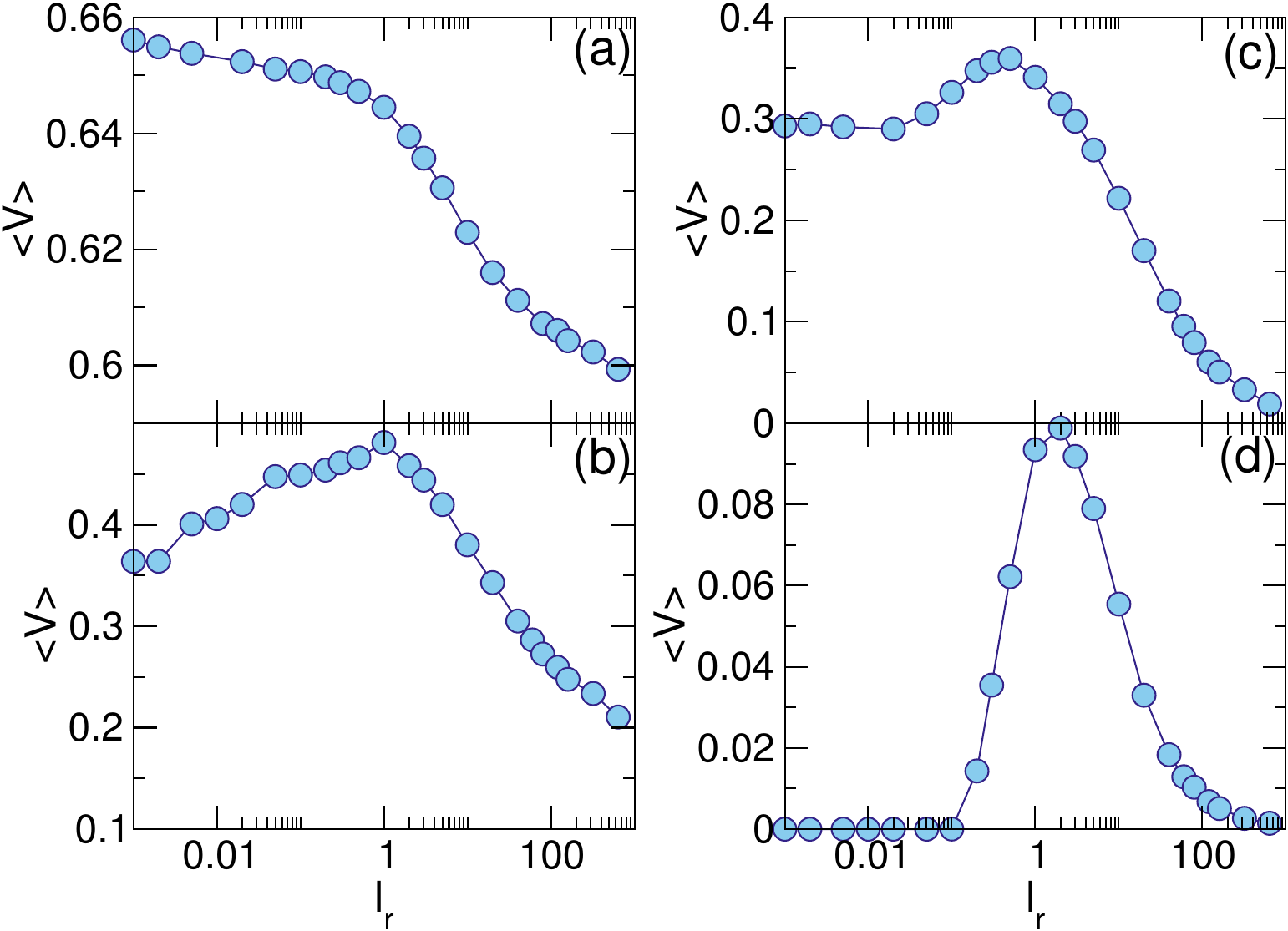}
\caption{ $\langle V\rangle$ vs $l_{r}$
  for samples with fixed $\phi_{\rm tot} = 0.157$ and varied $\phi_{\rm obs}$.
  (a) $\phi_{\rm obs}= 0.03146$.
  (b) $\phi_{\rm obs}=0.0628$.
  (c) $\phi_{\rm obs}=0.09424$.
  (d) $\phi_{\rm obs}=0.1257$.
  }
\label{fig:9}
\end{figure}

\subsection{Varied Obstacle Density and Motor Force}

We next consider the effect of holding
$\phi_{\rm tot}$ fixed at $\phi_{\rm tot}=0.157$ while decreasing 
$\phi_{\rm obs}$.
In Fig.~\ref{fig:9}(a) we plot
$\langle V\rangle$ versus $l_{r}$ for
a sample with $\phi_{\rm obs} = 0.03146$.
Here $\langle V\rangle$ decreases monotonically with increasing $l_{r}$ and the
flow persists 
even when $l_{r} = 0$.
At $\phi_{\rm obs}=0.0628$
in Fig.~\ref{fig:9}(b)
and
at $\phi_{\rm obs}=0.09424$
in Fig.~\ref{fig:9}(c),
there is still finite flow for
$l_{r} = 0$, and
a peak in $\langle V\rangle$ emerges
near $l_{r} = 1.0$.
At $\phi_{\rm obs} = 0.1257$ in Fig.~\ref{fig:9}(d),
$\langle V\rangle=0$ when $l_{r} < 0.1$, 
and the optimum flow,
indicated by the highest value of $\langle V\rangle$,
has shifted to a higher run length of
$l_{r} = 2.0$.

\begin{figure}
  \center
  \includegraphics[width=0.6\columnwidth]{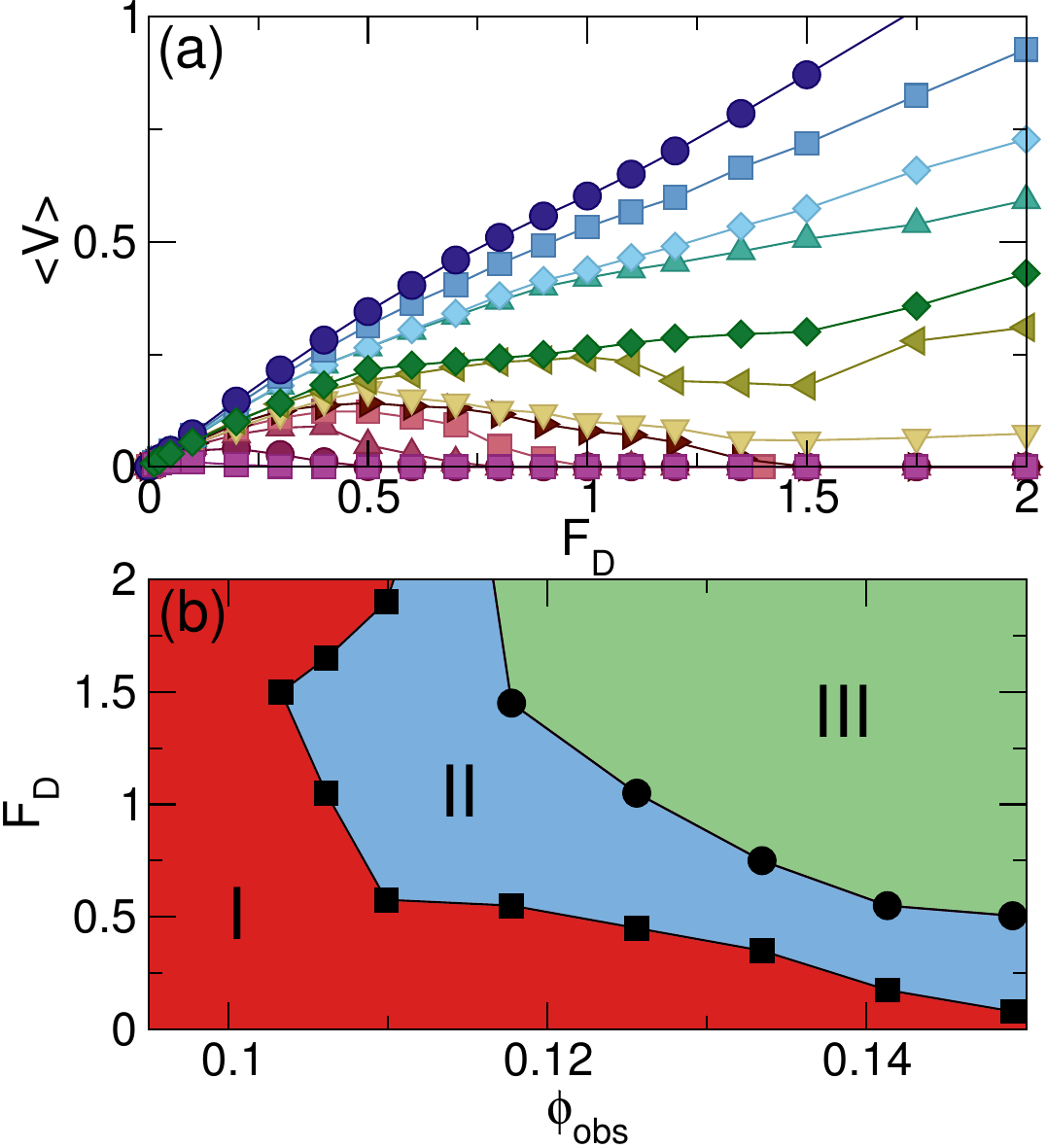}
  \caption{(a) $\langle V\rangle$ vs $F_{D}$ for
    samples with fixed $\phi_{\rm tot} = 0.157$ 
and varied $\phi_{\rm obs}$ at $l_{r} = 1.0$ and $F_m=0.5$. 
From top to bottom,
$\phi_{\rm obs} = 0.00785$
(dark blue circles),
$0.00864$
(medium blue squares),
$0.00942$
(light blue diamonds),
$0.102$
(teal up triangles),
$0.1033$
(dark green diamonds),
$0.10618$
(light green left triangles),
$0.1099$
(yellow down triangles),
$0.1178$
(dark red right triangles),
$0.1256$
(light pink squares), 
$0.133$
(dark pink up triangles),
$0.1413$
(dark magenta circles),
and $0.149$
(light magenta squares).   
(b) Dynamic phase diagram as a function of $F_D$ vs $\phi_{\rm obs}$ for the
system in panel (a), showing reentrance in phase I.
Phase I (red): ohmic flow; phase II (blue): partial trapping with NDM; phase III (green):
clogged.
}
\label{fig:10}
\end{figure}

In Fig.~\ref{fig:10}(a) we plot
$\langle V\rangle$ versus $F_{D}$ for a system with $l_r=1.0$, $F_m=0.5$,
fixed $\phi_{\rm tot} = 0.157$, and varied
obstacle density
ranging from
$\phi_{\rm obs}=0.007853$ to
$\phi_{\rm obs}=0.1492$.
The upper value of $F_D$ at which $\langle V\rangle$ drops to zero decreases
with increasing $\phi_{\rm obs}$ for
$\phi_{\rm obs} > 0.1178$, while 
for $0.1033 < \phi_{\rm obs} < 0.1178$ we observe a window of
NDM where $\langle V\rangle$ decreases with
increasing $F_{D}$ separating
low and high $F_D$ regions in which $\langle V\rangle$ increases
with increasing $F_D$.
For $\phi_{\rm obs} < 0.1033$,
$\langle V\rangle$ monotonically 
increases with increasing $F_{D}$, and although the NDM has disappeared,
there is still a
decrease in the slope of
$\langle V\rangle$
for $F_{D} > 0.5$
because an increased amount of trapping occurs once $F_D>F_m$.
Due to the harmonic form of the disk-disk interaction potential,
if $F_{D}$ is increased to a large enough value the disks eventually are able to
depin and move even in the completely clogged state; however, this occurs well
above the range of $F_D$ that we consider here.
In Fig.~\ref{fig:10}(b), the dynamic phase diagram as a function of $F_D$ versus
$\phi_{\rm obs}$ for the system in Fig.~\ref{fig:10}(a)
shows that the ohmic flow phase I is reentrant.

\begin{figure}
  \center
\includegraphics[width=0.8\columnwidth]{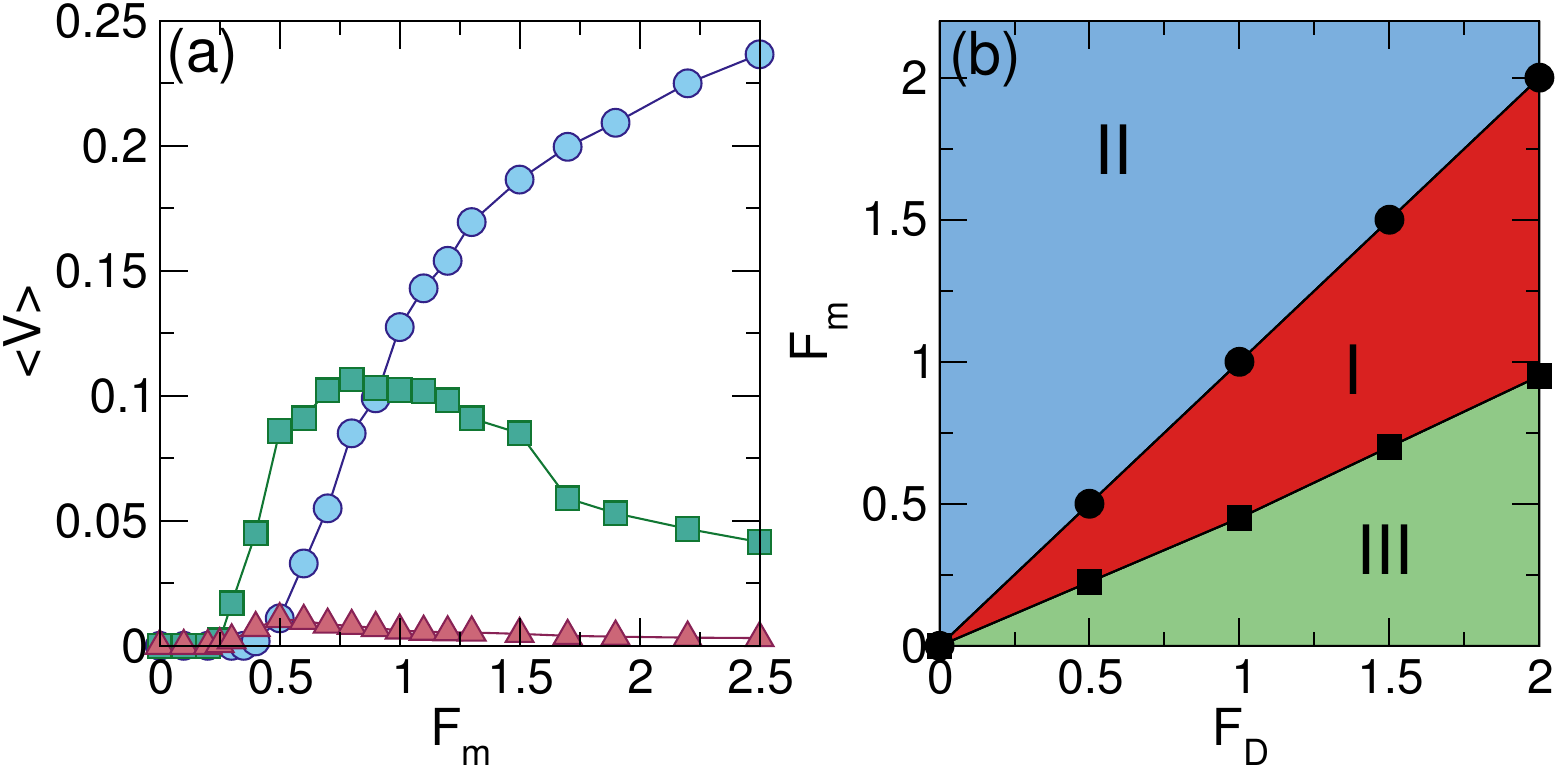}
\caption{
  $\langle V\rangle$ vs the motor force $F_{m}$
  for a system with $F_{D} = 0.5$,
  $\phi_{\rm obs} = 0.1257$, and $\phi_{a} = 0.0314$
  at $\tau =100$ (circles), $10000$ (squares),
  and $120000$ (triangles).
  (b) Dynamic phase diagram as a function of
  $F_{m}$ vs $F_{D}$ for the same system
at $\tau = 120000$.
}
\label{fig:11}
\end{figure}

Up to now we have characterized the activity
by the run length
$l_{r} = \tau F_{m}\delta t$ and have focused on the case
$F_{m} = 0.5$.
It is, however, possible to obtain 
different behaviors at fixed $l_{r}$
by varying $F_{m}$ and $\tau$. 
If $F_{D} < F_{m}$, the value of $\langle V\rangle$ should always be finite.
In Fig.~\ref{fig:11}(a) we plot
$\langle V\rangle$ versus $F_{m}$ in systems with fixed $F_{D} = 0.5$, 
$\phi_{a} = 0.0314$, and $\phi_{\rm obs} = 0.1257$ for three
values of $\tau$.  In order to compare these plots to our
previous results, note that for $F_m=0.5$, 
$\tau = 100$ gives
$l_r=0.1$,
$\tau = 10000$ is equivalent to
$l_{r} = 10.0$,
and $\tau = 120000$
corresponds to
$l_{r} = 120$.
For $\tau = 100$,
$\langle V\rangle = 0$ when $F_{m} < 0.4$, and for $F_m \geq 0.4$, $\langle V\rangle$
increases monotonically with $F_m$.
At $\tau = 10000$,
$\langle V\rangle > 0$ for $F_{m} > 0.2$  and $\langle V\rangle$ passes through a
maximum near $F_{m} = 0.8$.
For $\tau= 120000$,
the maximum in $\langle V\rangle$ falls at $F_{m} = 0.5$,
and the overall magnitude of $\langle V\rangle$ is much smaller than that observed
at the smaller $\tau$ values.
In Fig.~\ref{fig:11}(b) we plot a dynamic phase diagram
as a function of $F_m$ versus $F_D$ for the $\tau=120000$ system.
The I-III transition line separating the clogged phase III and the ohmic flow
phase I falls at
$F_{m} = F_{D}/2$,
while the NDM in phase II appears when
$F_{m} > F_{D}$ .

\begin{figure}
  \center
\includegraphics[width=0.5\columnwidth]{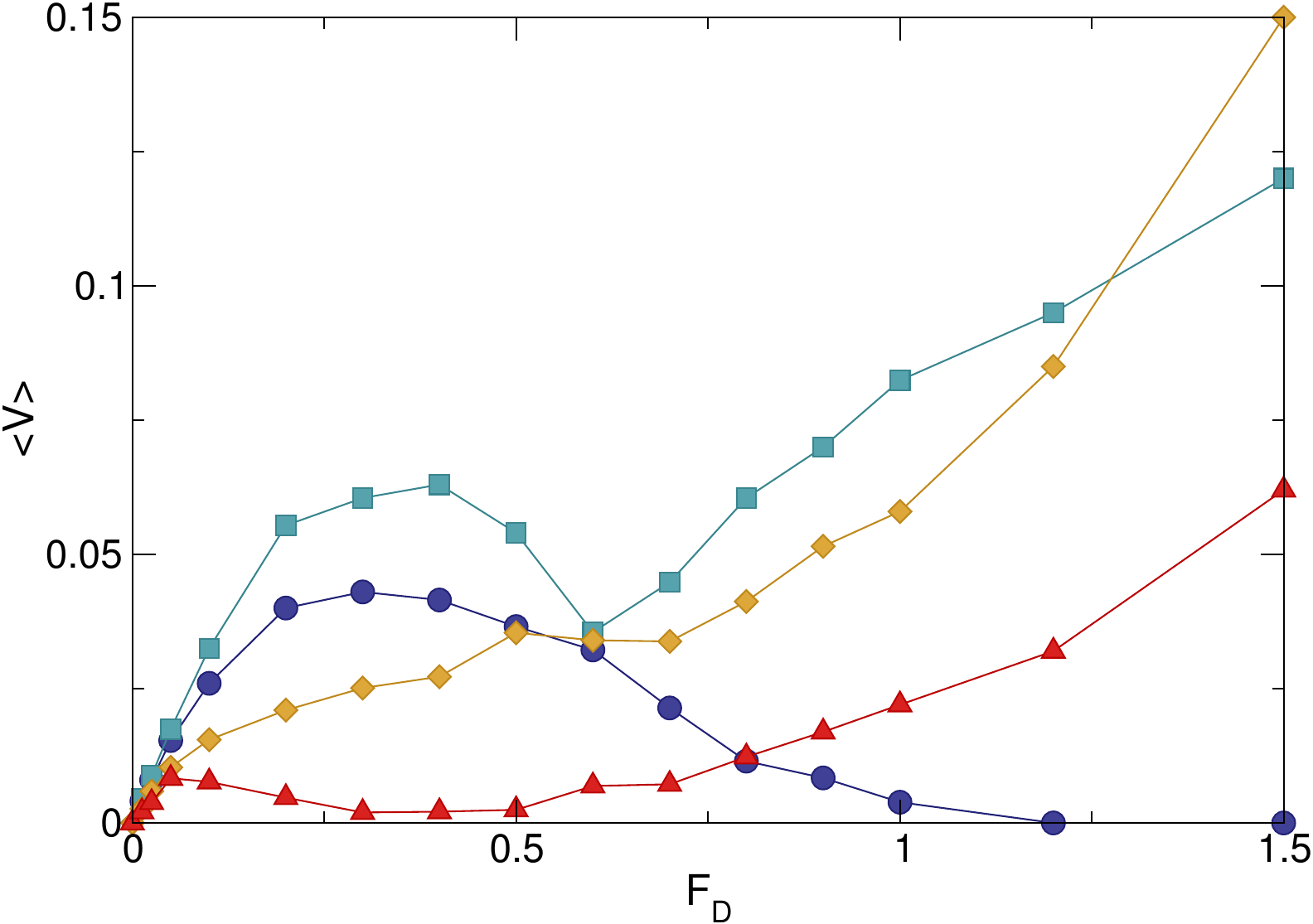}
\caption{ $\langle V\rangle$ vs $F_{D}$ for
  samples with $\phi_{\rm obs} = 0.173$,
$F_{m} = 0.5$, and $l_{r} = 1.0$ at $\phi_{\rm tot} = 0.25$ (blue circles), 
  $0.377$ (green squares), $0.565$ (orange diamonds),
  and $0.7$ (red triangles).
  }
\label{fig:12}
\end{figure}

In previous work examining the
mobility as a function of $\phi_{a}$
for fixed $\phi_{\rm obs}$ and fixed $F_{D}$, 
$\langle V\rangle$ increased with increasing $\phi_{a}$
up to a maximum value and then decreased for higher $\phi_a$ as the disks
approached the jamming density due to a
crowding effect
in which the active disks become so dense that
they impede each other's motion \cite{11}. 
In Fig.~\ref{fig:12} we plot $\langle V\rangle$
versus $F_{D}$ for a system with $\phi_{\rm obs} = 0.173$, $F_{m} = 0.5$,
and $l_{r} = 1.0$ at $\phi_{\rm tot} = 0.25$,
$0.377$,
$0.565$,
and $0.7$.
For $\phi_{\rm tot} = 0.25$, $\langle V\rangle$
drops to zero for $F_{D} > 1.25$, while
for $\phi_{\rm tot} = 0.377$,
there is a region of NDM for $0.4 < F_{D} < 0.7$ but the velocities remain finite
and the overall magnitude of $\langle V\rangle$ is larger than
that of the $\phi_{\rm tot}=0.25$ system.
For $\phi_{\rm tot} = 0.565$,
the average $\langle V\rangle$ is smaller
than that at $\phi_{\rm tot} = 0.377$ due to the crowding effect,
and there is only a very small window of NDM near $F_{D} = 0.6$.
For $\phi_{\rm tot} = 0.7$, the increased crowding effect causes
a substantial decrease in the overall magnitude of
$\langle V\rangle$, and at the same time an extended region of NDM
appears for
$0.05 < F_{D} < 0.4$. 

\begin{figure}
  \center
\includegraphics[width=0.6\columnwidth]{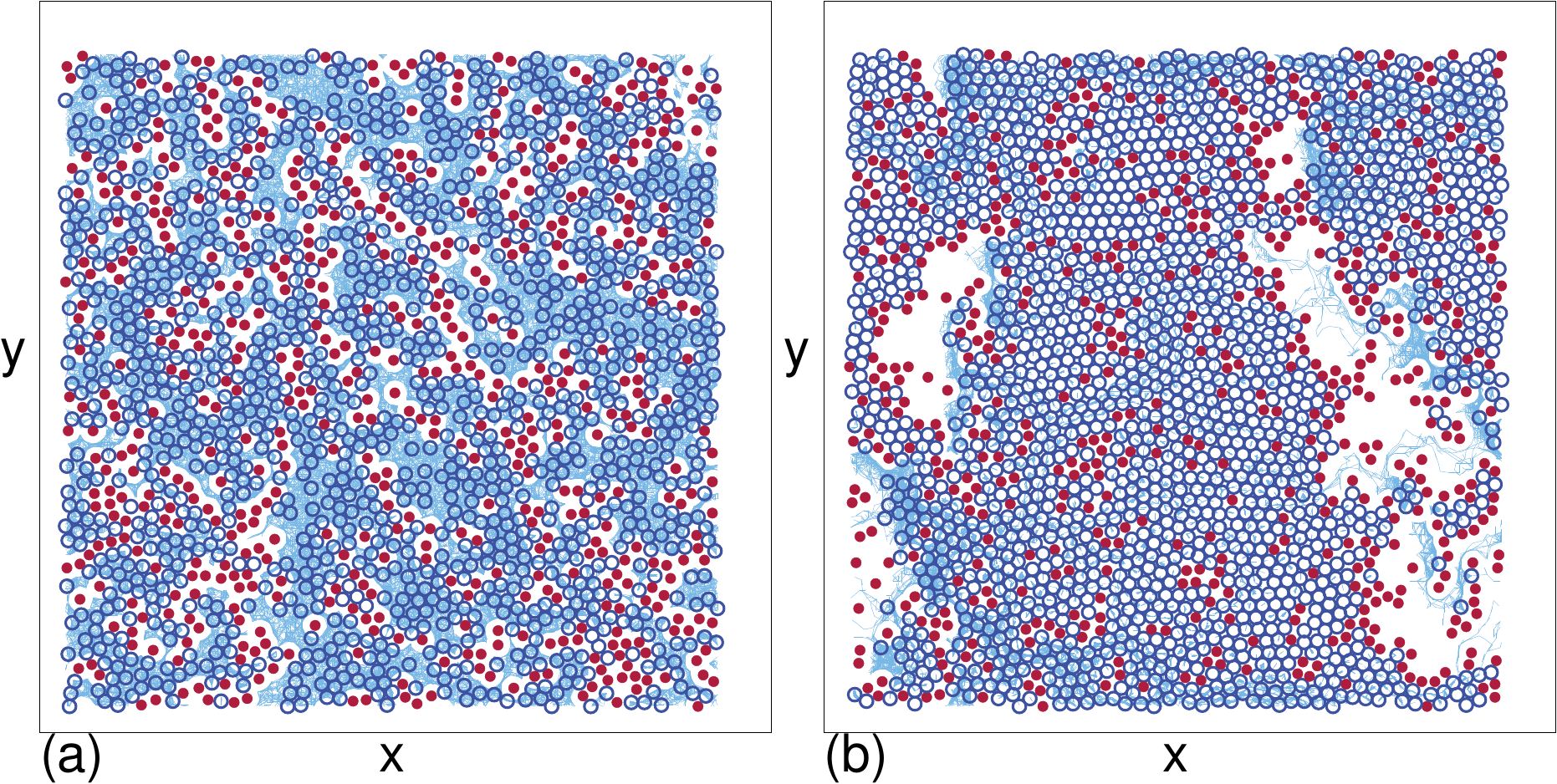}
\caption{ The obstacle positions (red filled circles), active disks
  (dark blue open circles), and trajectories (light blue lines)
  for the system in Fig.~\ref{fig:12} at (a) 
  $\phi_{\rm tot} = 0.565$ with $F_D=0.0125$ and (b) $\phi_{\rm tot} = 0.7$
  with $F_D=0.5$.
}
\label{fig:13}
\end{figure}

In Fig.~\ref{fig:13}(a) we plot the active disk trajectories
for the system in Fig.~\ref{fig:12}
at $\phi_{\rm tot} = 0.565$ and $F_{D} = 0.0125$,
a regime in which $\langle V\rangle$ increases with increasing $F_D$.
There is a considerable amount of disk motion throughout the system.
In contrast, Fig.~\ref{fig:12}(b) illustrates an NDM regime at
$\phi_{\rm tot} = 0.7$ and
$F_{D} = 0.5$, where
a large jammed or clogged area has formed
in the center of the sample, indicating the role played by crowding in
inhibiting the mobility of the active disks.

\section{Summary}
We have numerically examined the velocity-force curves for
active matter disks driven through  a random 
obstacle array and find three distinct dynamical phases.
In the low drive regime,
the velocity increases linearly
with increasing external drive.
For intermediate drives,
the system exhibits negative differential mobility 
where the velocity decreases with
increasing drive due to the trapping of disks
behind obstacles.
Finally, at high drive we find a fully clogged state in which the
drift velocity drops to zero.
For increasing activity or run length, we find that the onsets
of the NDM phase and the fully clogged phase are shifted to larger
external drift forces.
Additionally,
the drift velocity at fixed drive
changes nonmonotonically with increasing activity, indicating that there
is a drive-dependent optimal activity or run length that
maximizes the flux of disks through the system.
We map the locations of the dynamic phases
as a function of activity, active disk density, obstacle density, and motor force.
We describe how an external drift force could be  
tuned to either separate or mix active disk species with different mobilities.
We have also examined the
role of active disk density,
and find that at low disk densities, the NDM and clogging effects disappear
with increasing disk density when the trapping is reduced;
however, for much larger densities where crowding effects become important, the NDM
reappears and is enhanced.    

\ack
This work was carried out under the auspices of the 
NNSA of the U.S. DoE at LANL under Contract No.
DE-AC52-06NA25396.


\section*{References}
\begin{thebibliography}{99}

\bibitem{1}
  Marchetti M C, Joanny J F, Ramaswamy S, Liverpool T B, Prost J, Rao M and
  Simha R A 2013
 Hydrodynamics of soft active matter
  {\it Rev. Mod. Phys.} {\bf 85} 1143

\bibitem{2}
  Bechinger C, Di Leonardo R, L{\" o}wen H, Reichhardt C, Volpe G and Volpe G 2016
  Active Brownian particles in complex and crowded environments
  {\it Rev. Mod. Phys.} {\bf 88} 045006

\bibitem{3}
  Reichhardt C and Reichhardt C J O 2015
 Active microrheology in active matter systems: Mobility, intermittency, and avalanches
{\it Phys. Rev. E} {\bf 91} 032313

\bibitem{N}
  Fily Y and Marchetti M C 2012
  Athermal phase separation of self-propelled particles with no alignment
  {\it Phys. Rev. Lett.} {\bf 108} 235702

\bibitem{4}
  Redner G S, Hagan M F and Baskaran A 2013
  Structure and dynamics of a phase-separating active colloidal fluid
  {\it Phys. Rev. Lett.} {\bf 110} 055701

\bibitem{5}
  Palacci J, Sacanna S, Steinberg A P, Pine D J and Chaikin P M 2013
  Living crystals of light-activated colloidal surfers
  {\it Science} {\bf 339} 936

\bibitem{6}
  Cates M E and Tailleur J 2013
  When are active Brownian particles and run-and-tumble particles equivalent?
  Consequences for motility-induced phase separation
  {\it Europhys. Lett.} {\bf 101} 20010

\bibitem{7}
  Cates M E and Tailleur J 2015
  Motility-induced phase separation
  {\it Annu. Rev. Condens. Mat. Phys.} {\bf 6} 219

\bibitem{8}
  Thompson A G, Tailleur J, Cates M E and Blythe R A 2011
  Lattice models of nonequilibrium bacterial dynamics
  {\it J. Stat. Mech.} {\bf 2011} P02029

\bibitem{11}
  Reichhardt C and Reichhardt C J O 2014
Active matter transport and jamming on disordered landscapes
{\it Phys. Rev. E} {\bf 90} 012701

\bibitem{12}
  Morin A, Desreumaux N, Caussin J-B and Bartolo D 2017
   Distortion and destruction of colloidal flocks in disordered environments
  {\it Nature Phys.} {\bf 13} 63

\bibitem{13}
  Zeitz M, Wolff K and Stark H 2017
Active Brownian particles moving in a random Lorentz gas
{\it Eur. Phys. J. E} {\bf 40} 23

\bibitem{14}
  Lozano C, ten Hagen B, L{\" o}wen H and Bechinger C 2016
  Phototaxis of synthetic microswimmers in optical landscapes
  {\it Nature Commun.} {\bf 7} 12828

\bibitem{15}
  S{\' a}ndor Cs, Lib{\' a}l A, Reichhardt C and Reichhardt C J O 2017
  Collective transport for active matter run-and-tumble disk systems on a traveling-wave
  substrate
{\it Phys. Rev. E} {\bf 95} 012607

\bibitem{16}
  Choudhury U, Straube A V, Fischer P, Gibbs J G and H{\" o}fling F 2017
  Active colloidal propulsion over a crystalline surface
  {\it Preprint} arXiv:1707.05891

\bibitem{17}
  Pince E, Velu S K P, Callegari A, Elahi P, Gigan S, Volpe G and Volpe G 2016
  Disorder-mediated crowd control in an active matter system
  {\it Nature Commun.} {\bf 7} 10907

\bibitem{18}
  S{\' a}ndor Cs, Lib{\' a}l A, Reichhardt C and Reichhardt C J O 2017
Dynamic phases of active matter systems with quenched disorder
{\it Phys. Rev. E} {\bf 95} 032606 

\bibitem{N1}
  Reichhardt C J O and Reichhardt C 2017
Ratchet effects in active matter systems
{\it Annu. Rev. Condens. Mat. Phys.} {\bf 8}, 51

\bibitem{19}
  K{\" u}mmel F, Shabestari P, Lozano C, Volpe G and Bechinger C 2015
Formation, compression and surface melting of colloidal clusters by active particles
{\it Soft Matter} {\bf 11} 6187

\bibitem{20}
  Chepizhko O, Altmann E G and Peruani F 2013
  Optimal noise maximizes collective motion in heterogeneous media
  {\it Phys. Rev. Lett.} {\bf 110} 238101

\bibitem{21}
  Quint D and Gopinathan A 2015
   Topologically induced swarming phase transition on a 2D percolated lattice
  {\it Phys. Biol.} {\bf 12} 046008 

\bibitem{N2}
  Reichhardt C and Reichhardt C J O 2017
  Depinning and nonequilibrium dynamic phases of particle assemblies driven over
  random and ordered substrates: A review
  {\it Rep. Prog. Phys.} {\bf 80} 026501

\bibitem{22}
  Barma M and Dhar D 1983
Directed diffusion in a percolation network
{\it J. Phys. C} {\bf 16} 1451

\bibitem{23}
  Leitmann S and Franosch T 2013
Nonlinear response in the driven lattice Lorentz gas
{\it Phys. Rev. Lett.} {\bf 111} 190603

\bibitem{24}
  Baerts P, Basu U, Maes C and Safaverdi S 2013
Frenetic origin of negative differential response
{\it Phys. Rev. E} {\bf 88} 052109

\bibitem{25}
  B{\' e}nichou O, Illien P, Oshanin G, Sarracino A and Voituriez R 2014
Microscopic theory for negative differential mobility in crowded environments
{\it Phys. Rev. Lett.} {\bf 113} 268002 

\bibitem{26}
  Baiesi M, Stella A L and Vanderzande C 2015
Role of trapping and crowding as sources of negative differential mobility
{\it Phys. Rev. E} {\bf 92} 042121 

\bibitem{27}
  B{\' e}nichou O, Illien P, Oshanin G, Sarracino A and Voituriez R 2016
  Nonlinear response and emerging nonequilibrium microstructures for biased
  diffusion in confined crowded environments
{\it Phys. Rev. E} {\bf 93} 032128

\bibitem{28}
  Sarracino A, Cecconi F, Puglisi A and Vulpiani A 2016
  Nonlinear response of inertial tracers in steady laminar flows: differential and
  absolute negative mobility
{\it Phys. Rev. Lett.} {\bf 117} 174501

\bibitem{29}
  Jack R L, Kelsey D, Garrahan J P and Chandler D 2008
Negative differential mobility of weakly driven particles in models of glass formers
{\it Phys. Rev. E} {\bf 78} 011506 

\bibitem{30}
  Reichhardt C, Olson C J and Nori F 1998
  Nonequilibrium dynamic phases and plastic flow of driven vortex lattices in
  superconductors with periodic arrays of pinning sites
{\it Phys. Rev. B} {\bf 58} 6534

\bibitem{31}
  Gutierrez J, Silhanek A, Van de Vondel J, Gillijns W and Moshchalkov V 2009
  Transition from turbulent to nearly laminar vortex flow in superconductors with
  periodic pinning
{\it Phys. Rev. B} {\bf 80} 140514

\bibitem{33}
  Eichhorn R, Regtmeier J, Anselmetti D and Reimann P 2010
Negative mobility and sorting of colloidal particles
{\it Soft Matter} {\bf 6} 1858

\bibitem{34}
  Scholl E 1987
  {\it Nonequilibrium Phase Transitions in Semiconductors}
  (Berlin:Springer-Verlag)

\end{thebibliography}
\end{document}